# New Mechanism for Strongly Bound Excitons in Gapless Two-Dimensional Structures

Yufeng Liang[1], Ryan Soklaski[1], Shouting Huang[1], Matthew W. Graham[2], Robin Havener[3], Jiwoong Park[4, 5], Li Yang[1,*]

[1]*Department of Physics, Washington University in St. Louis, St. Louis, MO 63130, USA*
[2]*Laboratory for Atomic and Solid State Physics, Cornell University, Ithaca, NY 14853, USA*
[3]*School of Applied and Engineering Physics, Cornell University, Ithaca, NY 14853, USA*
[4]*Department of Chemistry and Chemical Biology, Cornell University, Ithaca, NY 14853, USA*
[5]*Kavli Institute at Cornell for Nanoscale Science, Cornell University, Ithaca, NY 14853, USA*

**Abstract:** Common wisdom asserts that bound excitons cannot form in high-dimensional (d>1) metallic structures because of their overwhelming screening and unavoidable resonance with nearby continuous bands. Strikingly, here we illustrate that this prevalent assumption is not quite true. A key ingredient that has been overlooked is that of viable decoherence that thwarts the formation of resonances. As an example of this general mechanism, we focus on an experimentally relevant material and predict bound excitons in twisted bilayer graphene, which is a two-dimensional gapless structure exhibiting metallic screening. The binding energies calculated by first-principles simulations are surprisingly large. The low-energy effective model reveals that these bound states are produced by a unique destructive coherence between two alike subband resonant excitons. In particular, this destructive coherent effect is not sensitive to the screening and dimensionality, and hence may persist as a general mechanism for creating bound excitons in various metallic structures, opening the door for excitonic applications based on metallic structures.

Bound excitons, electron-hole (*e-h*) pairs, are of particular interest because of their neat physics picture and intrinsic long lifetime that makes broad applications, including photovoltaic and photocatalytics [1-3]. However, the formation of bound *e-h* pairs had been thought to be impossible in metallic (gapless) systems due to their overwhelming screening effects. Moreover, *e-h* pairs in gapless structures tend to hybridize with continuous transitions nearby, forming resonant states, whose intrinsic lifetime is substantially shorter. To date, the only exception was found in metallic carbon nanotubes (mCNTs), in which the depressed one-dimensional (1D) screening together with the unique optical symmetry gap lead to the formation of a bound *e-h* pair [4-8]. Meanwhile, these studies ignite many obvious but fundamental questions: besides 1D metals, can we observe bound excitons in structures with stronger dielectric screening, *e.g.*, higher dimensional (d>2) gapless materials? In addition to the symmetry-related reason revealed in mCNTs, are there any other general mechanisms responsible for bound exciton formation in gapless systems?

Graphene, as a 2D semimetal, may serve as an excellent testbed to answer these outstanding questions. Unfortunately, due to a broad Fano resonance [9-12], no evidence of bound excitons has been observed, despite the presence of significant *e-h* interactions in graphene. Recently, twisted bilayer graphene (tBLG) [13-22], a 2D semimetal, has ignited substantial interest on their optical properties since a twist between graphene sheets introduces new van Hove singularities (vHSs) [13, 16, 17] that emerge at the intersections of Dirac cones on opposite layers. From the perspective of excitons, this unique band structure with several vHSs (see Fig. 1) has a particular implication for unusual excitonic effects. As shown in Figs. 1 (b) and (c), the outlined bands in each schematic are parallel to each other, due to the proximate group velocities of electrons and holes, which lead to a large joint density of states (JDOS). This special band topology enhances *e-h* interactions and therefore sheds new light on the potential existence of bound *e-h* pairs in 2D metallic systems.

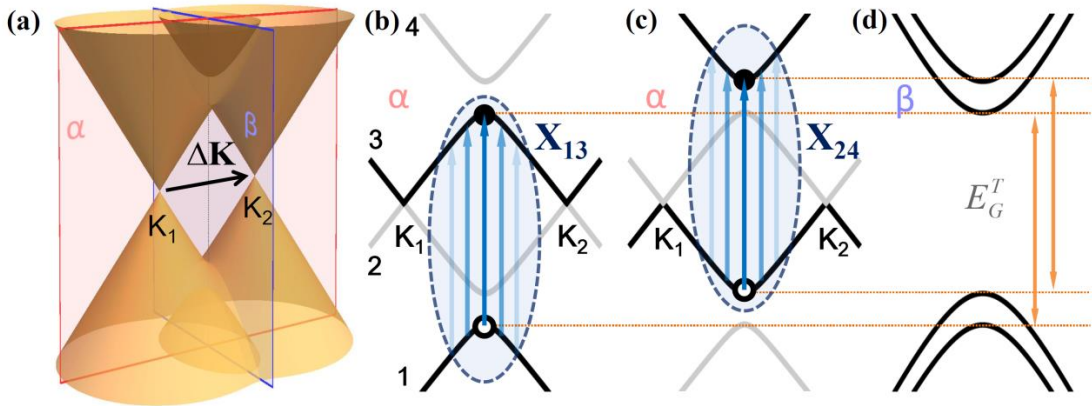

**FIG. 1** (color online). **(a)** Low-energy band structure of tBLG. α (light red) is the plane passing both axes of the Dirac cones whereas β (light blue) is the bisector plane of the two cones. **(b)(c)** Schematic formation of exciton $X_{13}$**(b)** and $X_{24}$**(c)** illustrated on the α-plane. The energy bands are labeled with 1 to 4 in ascending energy order. The involved bands are outlined in black while the states that mainly compose the exciton are enclosed by the ellipses. **(d)** Bands plotted on the β-plane. $E_G^T = \hbar v_F |\Delta \mathbf{K}|$ is the transition energy gap between $1^{st}$ ($2^{nd}$) and $3^{rd}$ ($4^{th}$) band.

In this letter, we predict the existence of strongly bound excitons in higher-dimensional gapless structures by a new decoherent effect, the Ghost Fano resonance. As an example of realistic material, we focus on excitonic effects of tBLG. Through first-principles GW-Bethe Salpeter Equation (BSE) simulations, we successfully observe a bound (though less bright) exciton with a significant binding energy of 0.5 eV in tBLG, which is an order of magnitude larger than that found in mCNTs [3-6] and is even comparable to those in semiconducting nanostructures [4-5, 23-27]. With the help of the low-energy effective model, we found that the formation of this unusual bound exciton is explained by the ghost Fano resonance [28, 29], a unique destructive coherence between two sets of resonant states with similar energies. This represents a new mechanism for forming bound excitons in gapless systems. In particular, because of its coherent origin, our proposed mechanism gives hope to creating bound excitons in many other metallic systems, despite their strong screening.

We perform first-principles calculations by employing the many-body Green's function theory for tBLG. We focus on two commensurate structures [15] with 21.8° and 32.2° rotated from the AB-stacking order. Our study begins with a density-functional-theory (DFT) calculation within the local density approximation (LDA) [30]. Next, the dielectric function is calculated using the random-phase approximation with a 30×30×1(18×18×1) $k$-grid [31] over the 1$^{st}$ Brillouin zone. Meanwhile a slab-Coulomb-truncation scheme [32] is also employed. We then obtain the quasiparticle (QP) band energies within the $G_0W_0$ approximation [33]. The vital step in describing the many-body excitonic effects is to solve the BSE [34].

$$(E_{c\mathbf{k}} - E_{v\mathbf{k}})A^S_{vc\mathbf{k}} + \sum_{v'c'\mathbf{k}'}\langle vc\mathbf{k}|K^{eh}|v'c'\mathbf{k}'\rangle A^S_{v'c'\mathbf{k}'} = \Omega^S A^S_{vc\mathbf{k}} \quad (1)$$

where $A^S_{vc\mathbf{k}}$ is the exciton wavefunction in k-space, $\Omega^S$ is the exciton eigenenergy, $K^{eh}$ is the $e$-$h$ interaction kernel and $|v\mathbf{k}\rangle$ and $|c\mathbf{k}\rangle$ are the hole and electron states respectively [34]. To ensure a smooth and accurate optical spectrum, we incorporated a fine 60×60×1 (36×36×1) $k$-grid in solving the BSE. Seven (Twelve) valence bands and seven (twelve) conduction bands are included to cover a broad range of the optical absorption spectrum up to 6.0eV.

Both optical spectra with and without $e$-$h$ interactions are presented in Figs. 2 (a) and (b) with three distinct peaks (marked by $E_1$, $E_2$, and $E_3$ in non-interacting spectra). Our calculation yields an excellent agreement with our recent optical conductivity measurements [19, 20]; the first two peaks, $E_1$ and $E_2$, stem from the two intersections between the Dirac cones from opposite layers, and the third one, $E_3$, results from the perturbed saddle-point vHSs intrinsic to monolayer graphene [18, 19]. We observe enhanced excitonic effects in the absorbance. $e$-$h$ interactions cause peaks $E_1$ and $E_2$ to redshift by ~0.2eV for both twist angles.

The fundamental mechanism forming the corresponding excitonic states in these new prominent peaks in tBLG ($E_1$ and $E_2$), however, may be substantially different from our knowledge learned from usual BLG [11, 35]. At the band intersection between two Dirac cones, only two sets of optical transitions with similar energies are allowed due to the selection rule, as shown in Fig.

1(b) and (c), producing double resonance [16, 17]. From the point view of two-particle excitations, the parallel sets of bands give rise to significant JDOS and potentially unusual bound *e-h* pairs.

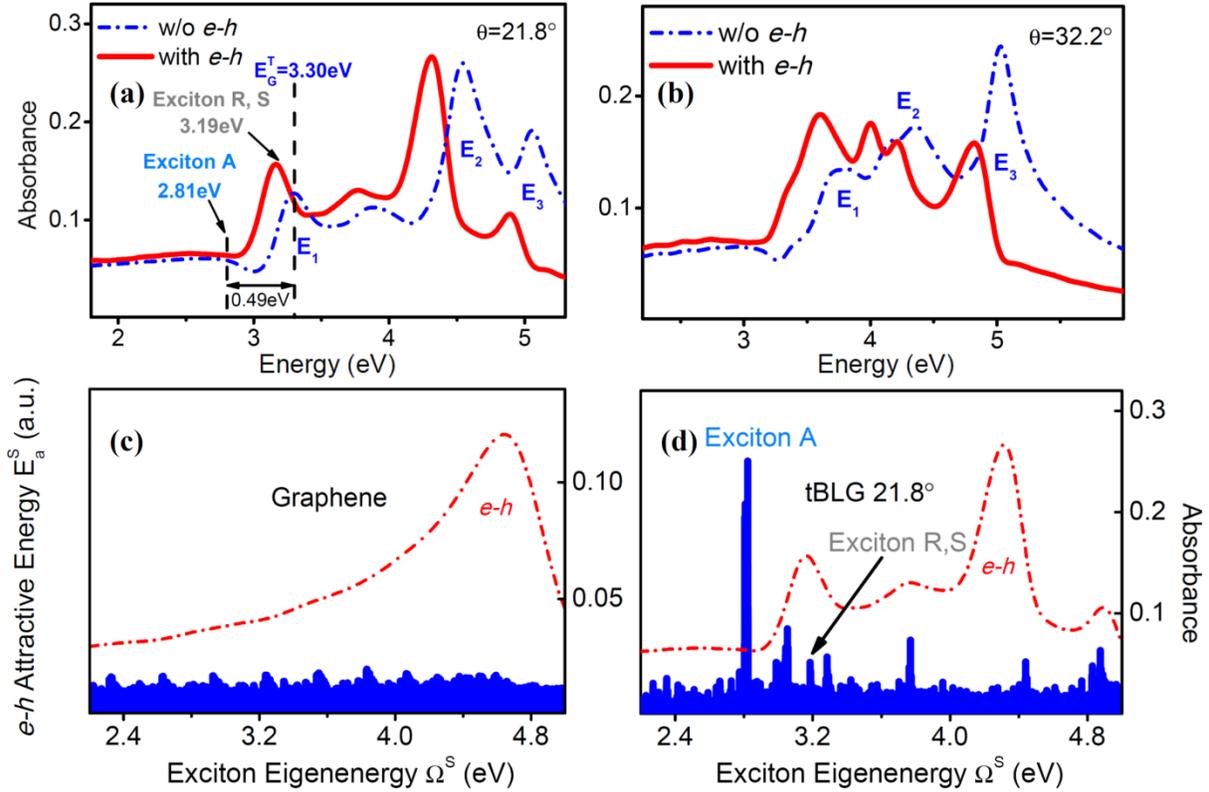

**FIG. 2** (color online). **(a)(b)** Optical absorbance obtained by the GW+BSE method. The blue dash-dotted curves are the non-interacting spectra while the red solid ones are the spectra with the *e-h* interactions included. **(c)(d)** *e-h* attractive energy $E_a^S$ (blue bars, in arbitrary unit) plotted versus the exciton energy $\Omega^S$ for graphene **(c)** and tBLG **(d)** within an identical energy window from 2.2-5.0eV. For references, the absorbance spectra of both structures are also plotted (red dashed curves).

The most direct approach to examine whether an excitonic state is bound or resonant is to investigate its wave function in real space. We plot the wave functions of two typical bright excitons, R and S, located around peak $E_1$ (marked in Fig. 2(a)). Here, R is the brightest excitonic states around the absorption peak. However, as shown in Figs. 3 (a) and (b), the electron is distributed loosely around the hole and even extends beyond our simulation range. These wave functions manifest a signature of resonant states, as observed in graphene [11] and CNTs [4-6]; the binding feature of excitons is substantially weakened by hybridization with continuous Bloch states that are spatially periodic and extended. In conclusion, these prominent peaks in Figs. 2 are dominated by resonant excitons, instead of bound ones.

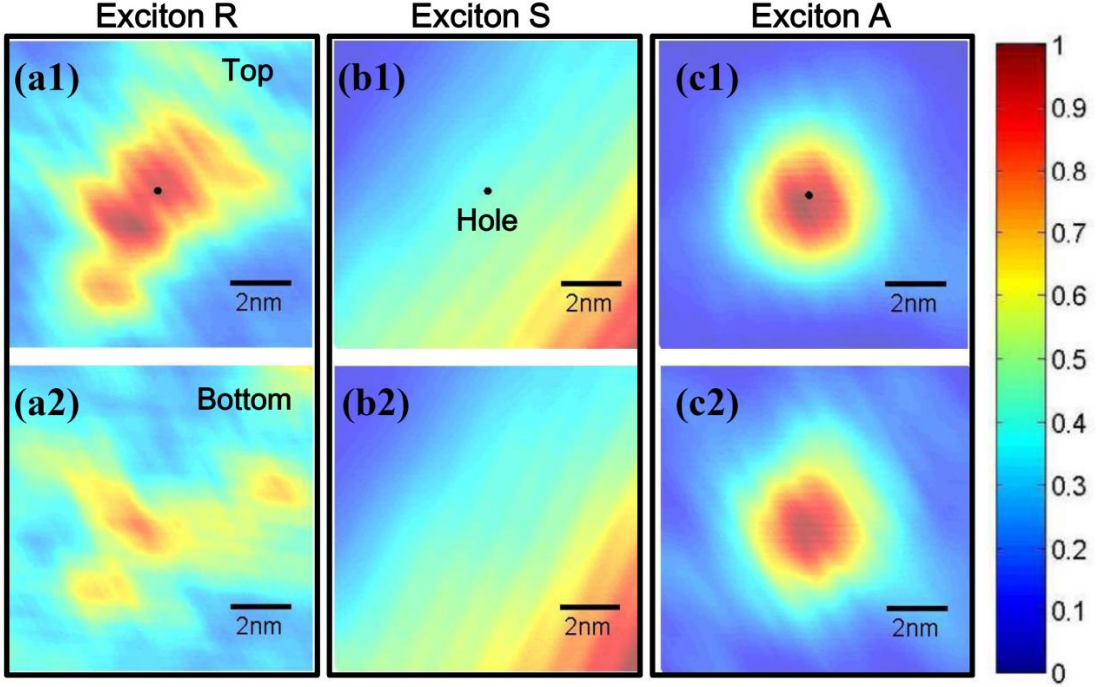

**FIG. 3** (color online) $|\chi_S(\mathbf{x}_e,\mathbf{x}_h)|^2$ of excitons R, S, and A in 21.8°-tBLG plotted on the top layer (a1)~(c1) and the bottom layer (a2)~(c2) with the hole fixed at the most-probable position on the top sheet. The distribution of the electron is normalized to the maximal probability of the two layers so that it ranges from 0 (blue) to 1 (red). The details within the primitive cell are less important and thus have been smoothened out.

So far we have focused on the brightest exciton, which often corresponds to the most bound state. However, bound states are not necessarily bright [36]. In order to find possible bound exciton states that are not optically active, we have to, in principle, scan all excitonic states solved by BSE and inspect their real-space wave functions, which is implausible because of a huge number of excitonic states (more than 170,000). Motivated by the fact that *e-h* interactions of bound excitons are typically more significant than those of resonant ones, we evaluate the *e-h* attractive energy for a given excitonic state S, by calculating $E_a^S$, expectation value of the *e-h* interaction kernel $K^{eh}$ sandwiched by that state $|S\rangle$:

$$E_a^S = \langle S|K^{eh}|S\rangle = \sum_{vc\mathbf{k}}(E_{c\mathbf{k}} - E_{v\mathbf{k}})|A_{vc\mathbf{k}}^S|^2 - \Omega^S \qquad (2).$$

$E_a^S$ is not the binding energy, but it can be understood as the difference between the exciton's "kinetic energy" and its eigenenergy, roughly reflecting the degree of *e-h* attractions.

Using this *e-h* attractive energy analysis, an intriguing comparison can be made between monolayer graphene and tBLG. For both cases, we plot the *e-h* attractive energy spectra ($E_a^S$ versus $\Omega^S$) for all exciton states in Fig. 2 (c) and (d). Surprisingly, the $E_a^S$ spectrum of graphene (see the blue-bar plot in Fig. 2 (c)) exhibits no distinct features up to 5.0eV, even for the prominent absorption peak at 4.6eV. This indicates that all its excitonic states are broadly resonant [11]; however, the $E_a^S$ spectrum of tBLG (Fig. 2 (d)) clearly shows several distinct spikes over a broad energy range, implying the existence of excitonic states with stronger *e-h* interactions. Following this idea, we select the most bound excitonic state, A (marked by an arrow in Fig. 2(d)), and plot its real-space wave function in Fig. 3(c1) and (c2). For this case, we obtain an isotropic distribution with significant localization. For the first time, our calculation predicts the presence of a bound exciton state in tBLG, a 2D gapless material.

More questions are raised regarding exciton A. First, its energy is not at the prominent absorption peak ($E_1$) but approximately 0.38eV below it. Moreover, its optical oscillator strength is weak, roughly one fifth of that of brightest excitonic state R. These are in conflict with the conventional wisdom; the most bound state is usually the most optically active one according to the hydrogen model. Second, since the position of the peak $E_1$ in the non-interacting spectrum indicates the transition energy $E_G^T$ between the valence and conduction vHSs, the bound exciton A emerges 0.49eV below the $E_G^T$ in Fig. 2(a). Such a surprisingly large binding energy (~ 0.5 eV) is an order of magnitude larger than that found in mCNTs [4-6] and it is even comparable to those exciton binding energies of semiconducting nanostructures [4-5, 23-27].

Unfortunately, it is challenging to directly analyze the results of our above first-principles simulation. Here, we use a low-energy effective model [22] for simplifying the analysis:

$$H(\mathbf{k}) = \begin{pmatrix} H_0(\mathbf{k},0) & T^+ \\ T & H_0(\mathbf{k}-\Delta\mathbf{K},\theta) \end{pmatrix}, \quad (3)$$

where the intralayer dispersion and the interlayer interaction are respectively:

$$H_0(\mathbf{k},\theta) = \hbar v_F \begin{pmatrix} 0 & e^{-i\theta}(k_x - ik_y) \\ e^{i\theta}(k_x + ik_y) & 0 \end{pmatrix}, \quad T = \Delta \begin{pmatrix} 0 & 1 \\ 1 & 0 \end{pmatrix},$$

The matrix *T* describes the average interlayer interaction between AB and BA stacking order, where $\Delta$ is the interlayer coupling strength. We approximate the screened Coulomb interaction in the direct term $K_d^{eh}$ with the 2D Coulomb potential $v_c(q) \propto 1/q$ but drop the exchange term $K_x^{eh}$ because of its lessened importance in the graphene-related systems [37]. With the model, we then solve the BSE on a uniform $k$-grid with approximately 2000 k-points in proximity of the two Dirac cones. As an example, we choose a tBLG with 5°-rotation with interlayer coupling strength $\Delta$ of 130meV.

Following the analysis via Eq. (2), we scan the *e-h* attractive energy spectrum obtained by our model BSE calculations. Now, the transition energy gap $E_G^T$ is 1.05 eV and we focus on the energy regime below it. Interestingly, as displayed in Fig. 4 (a), a series of discrete excitonic states $X_n (n=1,2,\cdots)$ are found with distinct *e-h* attractive energies $E_a^S$ alongside a background of resonant excitons (marked by grey bars). With ascending exciton energy, their population becomes denser towards $E_G^T$ whereas $E_a^S$ decreases monotonically. If $E_G^T$ and $E_a^S$ are regarded as a "band gap" and "binding energies", respectively, they exhibit standard features of bound excitons in semiconductors. In particular, we have plotted wave functions of the lowest few states $X_n$ in the reciprocal space, as shown in Fig. 4(b). We immediately see their bound-state nature. For example, the distribution of $X_1$ is highly analogous to 1s state of hydrogen. Also given the fact that $X_1$ possesses the largest *e-h* attractive energy, we can conclude $X_1$ corresponds to exciton A in our first-principles simulations.

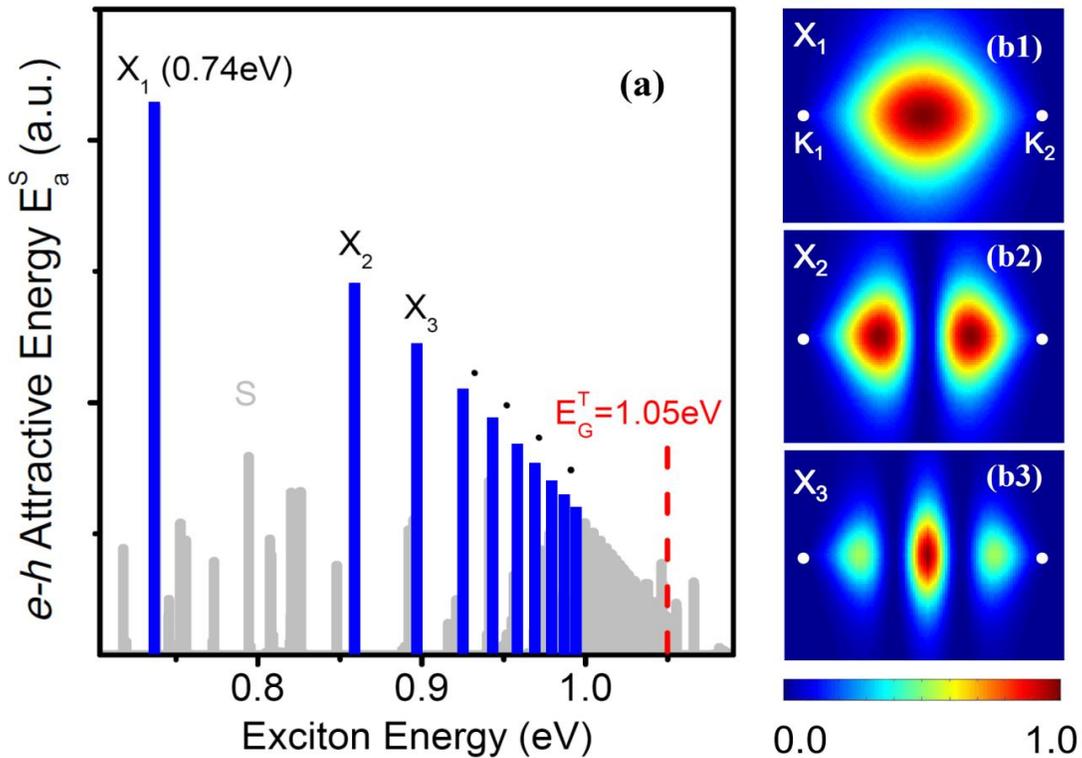

**FIG. 4** (color online). (a) *e-h* attractive energy from the low-energy effective model. The blue bars mark the exciton states with prominent *e-h* attractive energies whereas the gray bars represent the background states of less interest. (b1)~(b3) Modulus squared wavefunction of exciton $X_1, X_2,$ and $X_3$.

To further explain why we have bound excitons, $X_n$, in tBLG, we investigate their origins. Strikingly, we find that each state $X_n$ is only composed of two branches of double-resonant transitions (1→3 and 2→4), indicating that it is free of resonance with the Dirac continuum transition (2→3) occurring at low energies.

The above observation also inspires us to further investigate the relation of $X_n$ to the excitonic states solved on the 1→3 and 2→4 transition subspaces, for which we respectively obtain a set of subband bound excitons $X_{13,n}$ (Fig. 1(b)) and $X_{24,n}$ (Fig. 1(c)), emerging at identical energies $\Omega^{X_{13,n}} = \Omega^{X_{24,n}}$. Surprisingly, for each n, we find that the state $X_n$ are in fact purely antisymmetric superposition of two subspace excitons $X_{13,n}$ and $X_{24,n}$

$$|X_n\rangle = \frac{1}{\sqrt{2}}\left(|X_{13,n}\rangle - |X_{24,n}\rangle\right) \tag{4}$$

Because of the anti-phase coherence, the optical oscillator strength of $X_n$ is diminished. On the other hand, the symmetrically superposed states between $X_{13,n}$ and $X_{24,n}$ contribute to a set of higher-energy states, which are resonant and bright excitons. This understanding can be evidenced by Fig. 5(a), in which we present the projected density of states (PDOS) of the subband exciton $X_{13,1}$ (or $X_{24,1}$) over the full space $\{X_f\}$. Both $X_{13,1}$ and $X_{24,1}$ found near 0.78eV have 50%-overlap with $X_1$ occuring around 0.74eV, which is seen as a single spike in the PDOS. Meanwhile, they overlaps with a number of excitonic states at higher energy (around $\Omega^1 = 0.82$ eV), suggesting they have resonant components there. Moreover, although the oscillator strengths of $X_{13,n}$ and $X_{24,n}$ are individually bright, the destructive interference of the two components in exciton A renders its net oscillator strength relatively weak compared to the optically active higher-energy excitons, such as R and S.

In summary, the model calculation provides a surprising picture of excitonic interference as displayed in Fig. 5(b), in analogy with the so-called Ghost Fano resonance discovered in the model of quantum dot molecules [28, 29], in which the coherent effects between two similar-energy fano resonances give rise to a non-resonant energy level. First, although subband excitons $X_{13,n}$ and $X_{24,n}$ might hybridize with those 1→4 and 2→3 transition continua, they are also subject to mutual hybridization and are thrown into a symmetric state and an antisymmetric one. In the symmetric state, the coupling of $X_{13,n}$ and $X_{24,n}$ with the two transition continua interfere constructively, broadening into a group of bright excitonic states at higher energies via a conventional Fano resonance. Meanwhile, in the antisymmetric state, the couplings with the two continua cancel each other exactly, resulting in a dark and localized state $X_n$ at lower energy via the so-called Ghost Fano resonance [28].

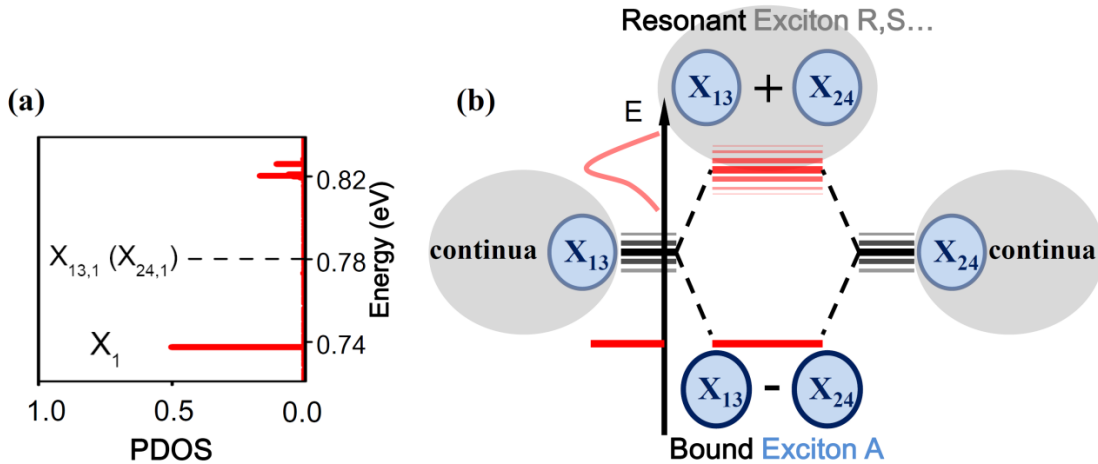

**FIG. 5** (color online). (a) PDOS $|\langle X_{13,1}|X_f\rangle|^2$ ($|\langle X_{24,1}|X_f\rangle|^2$) where $X_f$ goes over the full exciton space. (b) Exciton hybridization diagram in tBLG. The outlined circles represent the excitons formed on either the 1→3 or 2→4 transition subspaces while the grey ellipses represent the continua. The plus (minus) sign indicates the symmetric (anti-symmetric) superposition of exciton states.

We note that the above model is appropriate for small twist angle, and it may not be fully compatible with the quantitative results of our first-principles simulation, in which the twist angles are large (21.8° and 32.2°). However, the essential physics, such as the double resonance of transitions and the related destructive interference should still play an important role in shaping the strongly bound exciton A, even though the imperfect symmetry of conduction and valence bands could weaken the deconstructive effect, making exciton A not completely dark and not perfectly bound.

Finally, our predicted bound excitons will result in important experimental observations. Two-photon techniques or applying magnetic field may provide a means to detect them, as what had been done to observe dark excitons in CNTs [38, 39]. In particular, because the double-resonant picture holds better for tBLG with small twist angles due to the better *e-h* band symmetry, we expect the lifetime of optical excitations in tBLG will be much longer than that in graphene and BLG; the corresponding decay of optical currents shall become longer as the twist angle is reduced. More importantly, this formation mechanism is a coherent effect that is not strongly affected by the screening and *e-h* interaction strength. Therefore, we expect this phenomenon to be robust in many metallic systems.

In conclusion, we have demonstrated a novel mechanism for the formation of strongly bound excitons in 2D (semi-) metallic nanostructures via the decoehrent effect, the Ghost Fano resonance. A strongly bound exciton with a 0.5eV binding energy is identified in tBLG, which is an order of magnitude than previous one identified in mCNTs. More importantly, our predicted mechanism for forming strongly bound excitons is not sensitive to the screening and dimensionality, and hence may persist as a general mechanism for creating bound excitons in metallic structures. Therefore, this gives rise to room-temperature excitonic applications based on 2D and even higher-dimensional metallic structure.

This work is supported by the National Science Foundation Grant No. DMR-1207141 and by the Air Force Office of Scientific Research (FA9550-09-1-0691 and FA9550-10-1-0410). The computational resources have been provided by the Lonestar of Teragrid at the Texas Advanced Computing Center (TACC). The DFT calculation is performed with the Quantum Espresso [40] while the GW-BSE calculation is done with the BerkelegyGW package [41].

* lyang@physics.wustl.edu